# Coded Divergent Waves for Fast Ultrasonic Imaging: Optimization and Comparative Performance Analysis

Yasin Kumru, and Hayrettin Köymen, Senior *Member, IEEE*

*Abstract*—In this paper, we present the optimal use of coded signals in diverging wave transmission for fast ultrasonic imaging. The performance of coded imaging with diverging waves, quantified by SNR, CNR, speckle power and target signal strength, is optimized in terms of code length, wave profile, aperture size and field of view. We obtained one frame in 200 microseconds using coded diverging wave transmission, equivalent to very high 5000 frames/s rate, where the imaging depth is 7.5 cm. The performances of coded diverging wave transmission and conventional single focused phased array imaging are compared on a single frame basis. Complementary Golay sequences with code lengths ranging from 2 to 10 bits are used to code the signal. The signal strength and SNR obtained using synthetic transmit aperture and conventional single focused phased array imaging techniques, respectively, are used as reference in the performance analysis. We present a method to determine the optimum diverging wave profile to maximize the SNR across the entire viewing sector, which matches the received signal strength distribution to that of synthetic transmit aperture imaging. Optimum diverging wave profiles are determined for different sector angles and for different array apertures. For 90° sector, the SNR of 8-bit coded signal with optimized diverging wave profile is higher than that of conventional single focused phased array imaging at all depths and regions except the focal region, where it is 2-8 dB lower.

## I. INTRODUCTION

ULTRASOUND imaging is easily accessible, noninvasive, nonionizing and a commonly used modality in medical imaging. As the use of ultrasound imaging expanded, high temporal resolution, i.e., high frame rate became an important demand for applications such as blood flow imaging, cardiovascular imaging, shear wave elasticity imaging, and 3D/4D real time imaging [1]. An important challenge is the compromise between temporal resolution, SNR, CNR, range resolution, and lateral resolution. Although progress have been made to increase the frame rate, ultrafast imaging with high SNR and CNR remains unresolved.

Frame rate mainly depends on the data acquisition time in pulse-echo ultrasound imaging, which is limited by the time of propagation of sound to the depth of interest and back. It also depends on the lateral resolution and the field of view. Limiting either the lateral resolution or the field of view reduces the number of transmit events needed per frame allowing higher frame rates. One such method was proposed in [2]. The RF data from human heart in vivo were acquired at 260 frames/second (fps) at the expense of decreased lateral resolution and field of view.

Conventional imaging techniques cannot meet the temporal resolution demand required for ultrafast imaging. In these techniques, image lines are formed sequentially, each of which is generated with different transmission events [3]. The need for multiple transmission events to form the entire image causes a significant decrease in frame rate, which limits the use of conventional methods for fast imaging. If a single transmit focus is used; SNR, CNR and resolution in the nonfocal regions are compromised. Using multiple transmit foci addresses that but at the expense of further decrease in frame rate [4], [5].

Different methods have been devised over the years. One of the most commonly employed method is synthetic transmit aperture (STA) imaging, a version of which is originally developed for SONAR applications [6]. STA imaging relies on combining low resolution images obtained from consecutive transmissions from an element at a time to form a high-resolution image. The resulting image is obtained by dynamically focusing in transmit and receive at every imaging point [7]. Although STA imaging offers high image resolution compared to conventional imaging techniques, the frame rate is inversely proportional to the number of transmission events. STA technique also suffers from poor SNR [8] due to single element excitation and motion artifacts [9] due to long time span required to complete the data acquisition from all transmit/receive element pairs. A suitable strategy to increase the SNR is to use larger transmit apertures during each transmission event. This strategy is first applied to circular aperture [10] and extended to plane aperture [11]. The approach in [11] is based on using subapertures excited with a time delay profile that generates a diverging wave (DW). The time delay profile calculation is based on the virtual point source position and inter-element pitch. It was experimentally shown that it offers better performance than standard STA imaging in terms of SNR and CNR and provides similar image quality compared

This work was supported by the Scientific and Technological Research Council of Turkey (TUBITAK) under project grant 119E509.

Y. Kumru and H. Köymen are with the Department of Electrical and Electronics Engineering, Bilkent University, Ankara, Turkey (e-mail: yasin@ee.bilkent.edu.tr).

to conventional phased array imaging. This approach provides 30 fps, which is comparable to conventional imaging.

The concept of fast imaging using unfocused transmit waves, such as DW [12], [13], [14] and plane wave (PW) [15], [16], has been extensively studied in the literature. A method for real time 3-D ultrasound imaging with DW and mechanically scanned arrays is investigated in [12]. In this method, diverging pulses using several groups of transducer elements were proposed. The simulation results demonstrated that images could be obtained at 1000 fps and SNR similar to that of conventional methods can be obtained only if the peak transmitted pressure is increased. Another fast-imaging technique using diverging beams was demonstrated in [13] that utilizes limited number of active transmit and receive elements. It was shown that diverging beam transmission combined with synthetic aperture technique using 8 transmit and 64 receive elements provides reasonable image quality and higher frame rates than those obtained in conventional linear array imaging. In addition, a method for motion velocity estimation was proposed in [14], which realizes 6250 fps with a single diverging beam transmission.

PW transmission was discussed for cardiac ultrasound imaging in [15], in which an overview of the technical principles behind imaging modalities was provided. It was shown that PW transmission is a promising method to avoid long acquisition times. Recently, the application of PW transmission to the field of flow imaging was introduced in [16], in which the combination of PW transmission with transverse oscillations was proposed enabling to reconstruct high frame rate images and 2-D vector flow maps.

The use of single unfocused wave has been extended to the use of several steered DWs with coherent spatial compounding technique [17], [18], [19]. Sector images of a human heart for 120 mm imaging depth were obtained with 900 fps at 2.7 MHz in [17] and 316 fps at 3.75 MHz in [18].The performance of compounded DWs for cardiac strain imaging is examined and 700 fps is achieved at 2.5 MHz and 110 mm imaging depth in [19]. Coherent spatial compounding of multiple steered PWs was also implemented for a linear array transducer in [20], [21]. It was experimentally demonstrated in [20] that 1000 fps at 60 mm imaging depth could be achieved with 4.5 MHz linear array transducer without sacrificing image quality. An optimization method for PW imaging was proposed in [21] that offers 238 fps for 4.1 MHz linear array transducer if the pulse repetition frequency is 5 kHz.

To increase the frame rate without compromising the SNR, another preferred approach is to use coded excitation [22], [23]. It was shown that, coded excitation is a good candidate for medical ultrasound imaging to achieve high frame rate and SNR. Only complementary Golay sequences (CGSs) provide side lobe cancellation among all coding sequences proposed in the literature for medical ultrasound imaging. However, they require two successive transmissions, which reduces the frame rate by 2 folds [24]. The potential of coded excitation with DW transmission is investigated in [25] for cardiac imaging. The experimental results demonstrated that 2500 fps is possible with CGS coded signals at 2 MHz and 150 mm imaging depth.

To overcome the frame rate reduction, mutually orthogonal CGSs were proposed to code the transmitted signals in [26]. Spatio-temporal encoding with two-pairs of orthogonal CGSs for DW compounding has been also investigated in [27]. It was shown that orthogonally coded transmission compensates the frame rate reduction and offers high quality cardiac imaging compared to uncoded scheme and conventional phased array imaging when an appropriate number of steering angles are used.

Various spatial coding schemes are proposed in the literature. One of them involves applying coding matrices, such as Hadamard matrix, to spatially encode the entire linear transducer array [28]. Hadamard encoded DW was proposed instead of single pulse DW compounding. The goal was to create a series of virtual sources with overlapping subapertures. Each virtual source is coded with Hadamard matrix and a short time delay is applied between the emissions of each virtual source. It was stated that, this method provides 500 fps when 8 virtual sources are created and offers SNR improvement compared to standard DW compounding method with the same frame rate.

The performance of coded signals in medical ultrasonic imaging is impaired by the attenuation in the tissue. The correlation between the received signal and the reference signal deteriorates at the deeper imaging regions. Also, using DW for fast imaging further complicates the problem since the transmitted waves from the array elements are, largely, incoherently superimposed in the medium. We studied the problem particularly in view of these two adverse effects. We present the optimal use of coded signals in DW transmission for fast ultrasonic imaging in this paper.

We used the CGSs to code the signal. We compared the performances of coded DW, STA and conventional single focused (CSF) phased array imaging schemes without apodization. We used SNR, CNR, speckle power, signal strength as performance metrics. We optimized the performance of coded transmission with DW in terms of code length, wave profile, aperture size and field of view. We show that it is possible to achieve 5000 fps by using coded DW transmission, without compromising SNR.

## II. METHOD

### A. Coded Transmission

Use of coded signals increases available energy at a given peak transmitted pressure without sacrificing bandwidth [29]. This property can be used to improve the SNR and related performance in ultrasound imaging.

Coded transmission requires correlation receivers for detection where correlator output is maximum at zero phase shift and it is low at other phase shifts [30]. These lower correlator outputs are referred to as code side lobes. The presence of code side lobes affects the range resolution. We used the term range lobe rather than side lobe to avoid confusion with the lateral side lobes.

The CGSs have the unique property that they offer zero code range lobe. A CGS consists of a pair of sequences, CGS(A) and CGS(B). The sum of the autocorrelation functions of two sequences doubles at zero phase shift while it is zero for non-zero phase shifts [31]. We used the regular implementation of the coded signal. CGSs of length 2, 4, 8 and 10 bits are used to code the transmitted signal with 2 cycles/chip. We compared the results of these coded signals with that of a 2-cycle pulse.

We used Binary Phase Shift Keying to modulate the coded signals. In coded signals, the symbol used for each code bit is referred to as chip. For example, the length of an 8-bit coded signal is 8 chip lengths.

The receiver for coded signals incorporates a matched filter (MF) determined by the coded transmit sequence, which can be implemented in various ways. We employed correlation receiver implementation, which is common in real-time applications [32]. Since matched filtering is a linear operation, it can be used directly on the received channel data as well as on the beamformed data, equivalently.

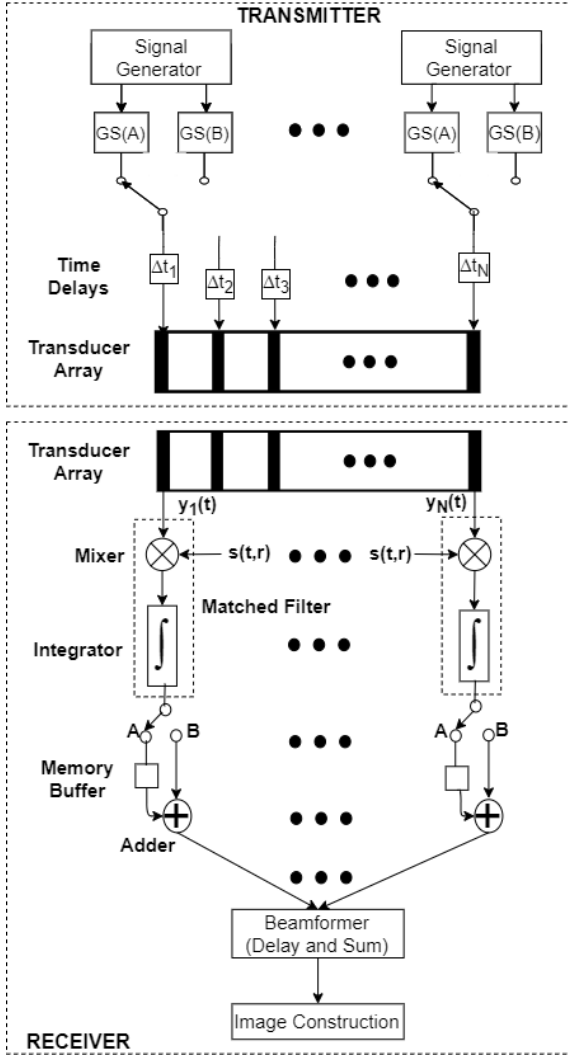

Fig. 1. Transmitter and receiver structure for coded excitation. GS(A) is a Golay sequence and GS(B) is its complementary sequence.

The transmitter and receiver block diagram used in this work for Golay coded excitation is shown in Fig.1. Every transducer array element is driven by CGS(A) and CGS(B) coded signals successively. Signals at each element are delayed to have a required wave profile. First, CGS(A) coded signals are transmitted from the transducer array and reflected waves are received by the same array. The received signals at each array element are filtered with a MF consisting of a mixer and integrator. The MF output for the $i^{th}$ transducer array element $\hat{R}_{ys,i}$, is expressed as:

$$\hat{R}_{ys,i}(m) = \sum_{k=0}^{K-1} y_i(m+k)\, s_r(k) \quad (1)$$

where $y_i(l)$ is the $l^{th}$ sample of the received signal at the $i^{th}$ transducer array element. $s_r(k)$ is the $k^{th}$ sample of the reference signal with length K. K is 54 for 1 chip and 203 for 8-chip signal. The reference signal is compensated for the attenuation in the medium and it is updated 12 times at every 0.5 cm depth. The depth index, $r$, refers to the updated reference signal, where $r$ ranges from 1 to 12. The compensation is described in detail in subsection B. The MF output, expressed in (1), for CGS(A) coded transmission is buffered until the next CGS(B) coded transmission data is collected. Then, the MF outputs at each array element for two transmissions are added. The resulting signals are then beamformed in delay and sum receive beamformer to form the image scan lines. We employed Hilbert transforming on the scan line data to obtain envelope information for image construction.

### B. Reference Signal in Attenuating-Medium

The correlation receiver requires a reference signal to compress the received signal. SNR is maximized if the received data contains the replicas of the reference signal [33]. Using the driving transmit waveform as the reference signal yields poor correlation, because both the transducer transfer function and the attenuation in the medium modifies this signal very significantly. We observed that the SNR in the correlator output deteriorates by more than 10 dB compared to using the reference signal extraction method described below in this section.

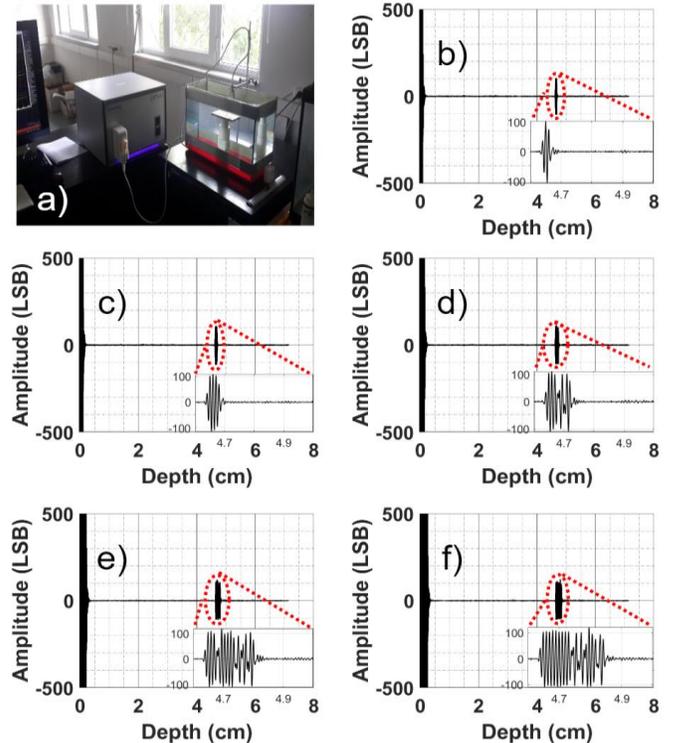

Fig. 2. The water measurement setup is illustrated in (a). The signals reflected from the iron plate and received by the 64$^{th}$ element of the phased array transducer are also shown. (b) 1 chip, CGS (A) coded signal with (c) 2, (d) 4, (e) 8, and (f) 10 chips.

We used two reference signals, one for each CGS, to decompress the coded received signal prior to beamforming. The reference signal must be two-way transmitted-and-received signal so that the effect of transducer transfer function is included in the reference signal. We performed a measurement in fresh water to obtain two-way transmitted-and-received signal. A highly reflecting material, a thick iron plate, was immersed into the water at 4 cm depth. The mid-element, 64th element, of the phased array transducer is fired and pulse echo measurements were conducted for all code lengths. Fig. 2 shows the measurement setup and received signals at the 64th element. Fresh water is almost attenuation-free compared to tissue and the signals preserve the bit-transitions when the code bits go from 0 to 1 or vice versa. This signal is an approximation to the reference signal, since the two-way electro-mechanical transfer functions of all combinations of transducer elements are not exactly similar.

The effect of attenuation on the complementary property of Golay pairs are discussed in [29] and it is reported that the attenuation prevents the perfect cancellation of code range lobes. This effect of the attenuation must be imposed also on the reference signal to improve the MF output in attenuating medium. The attenuation in the phantom used in this study (Model 550, Breast & Small Parts Phantom, ATS Laboratories, Bridgeport, USA), is specified as 0.5 dB/[MHz × cm]. We applied attenuation correction to the two-way transmitted-and-received signal at every 0.5 cm up to 6 cm depth. This correction is performed on the reference signal in frequency domain by applying exponential attenuation at respective depths. This correction provides an estimate of the attenuated reference signal. This attenuation compensated reference signal provides higher SNR compared to uncompensated two-way signal as the penetration depth increases. For example, the SNR obtained by the compensated reference signal at 3 cm depth is 3 dB higher.

We converted the reference signals to unit energy signals first and then normalized them such that the energy of each coded signal is proportional to the number of chips it contains. Thus, single chip reference signal has unit energy, the energy of two chip coded signal is 2, etc. This corresponds to keeping the amplitude of the coded signals fixed, regardless of the code length. We then used the resulting signals as a reference signal in correlation receiver. The effect of the attenuation correction for pulsed (1 chip) and CGS (A) coded signals with 2, 4, and 8 chips are shown in Fig. 3. It is apparent that the coded signals are significantly distorted as the depth increases regardless of their duration.

### C. Measurement Set-up

We collected the data using an ultrasound research scanner, Digital Phased Array System (DiPhAS, Fraunhofer IBMT, Frankfurt, Germany). The recorded raw data is sampled at 80 MHz. We used a phased array transducer (Fraunhofer IBMT, Frankfurt, Germany) operating at 7.5 MHz center frequency with a fractional bandwidth of 70%. There are 128 elements in the array, which are spaced by 0.1 mm pitch. The 2-cycle pulse and coded signals were produced as pulse width modulated signals before applying to the transducer for the transmission. The driving signal amplitude is kept constant. The phantom used in this study, Model 550, is designed with monofilament nylon line targets (pin targets), anechoic cyst structures and constructed of rubber-based material, structure of which is shown in Fig. 4. The monofilament nylon line targets (pin targets) have a diameter of 50 μm and have very weak reflectivity at 7.5 MHz. The rubber-based tissue-mimicking material has sound velocity of 1450 m/s ± %1 at 23ºC. The phantom temperature and the ambient temperature were recorded. We applied 22 dB fixed gain and 2.3 dB/cm linearly increasing time varying gain.

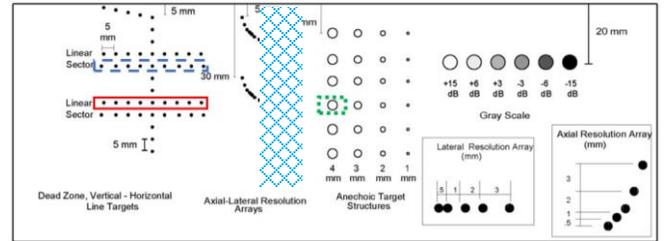

Fig. 4. The structure of the commercial ultrasound phantom used in the measurements.

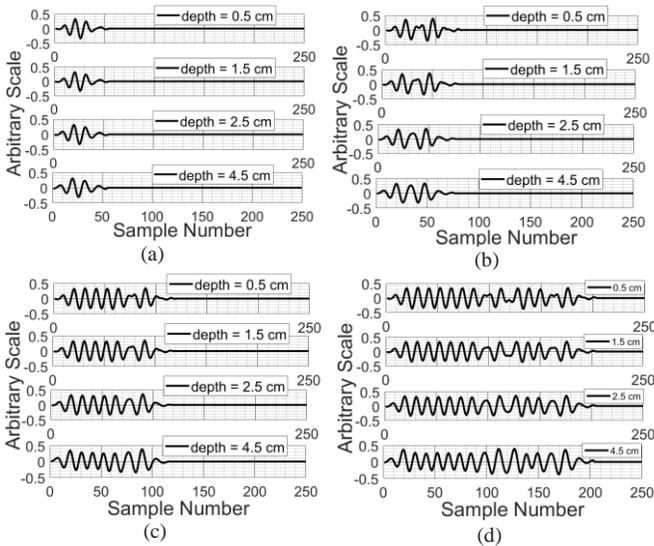

Fig. 3. Received signals on the 64th element of the phased array transducer after attenuation correction. (a) Pulsed with 1 chip, (b) CGS(A) coded signal with 2 chips. (c) CGS(A) coded signal with 4 chips. (d) CGS(A) coded signal with 8 chips. Every row within each figure corresponds to adaptive attenuation correction with respect to depths of 0.5, 1.5, 2.5, and 4.5 cm (return path length of 1, 3, 5, and 9 cm), respectively.

### III. DW Transmission for Fast Sector Imaging

#### A. Use of DW Transmission with Phased Array Transducers

The phased array transducers are used to produce sector images. They have smaller aperture but with numerous narrow-width elements spaced with about half-wavelength pitch. Conventional use of these transducers relies on multiple focused and steered beam transmissions yielding high resolution images typically at 25-30 fps.

PW transmission is not suitable for phased array transducers to increase the frame rate, since the covered field of view is very limited in the lateral dimension due to small transducer array aperture. One approach to increase the frame rate is using DW transmission, which insonifies the entire sector to be imaged.

DW profile is obtained by applying appropriate phases to array elements so that a cylindrically DW emanates from the array. The phase profile used for a particular DW is determined by the geometry given in Fig. 5, where the transmitted wave is assumed to be generated by an equivalent virtual source [18]. The assumed virtual source is a cylindrical omni directional line source, positioned behind the transducer array center at a distance of $r_v$. The omnidirectional cylindrical wave is assumed to be windowed by the array aperture and thus the insonification sector is established. The element pitch, $d$, is half of the wavelength at 7.5 MHz. The aperture length is $Nd$, where $N$ is 128. When the delay profile is configured for short $r_v$, the virtual source is close to the aperture and the total available energy is diverged to a wide insonification sector. When the profile is set for a long $r_v$, energy is confined to an acute sector. However, actual energy distribution in the sector differs from the geometrical predictions due to diffraction.

Keeping the transmitted ultrasound energy within the region of interest has a crucial role to increase the SNR in DW transmission. In this work, we present DW profile optimization for maximum SNR. We experimentally investigated the performance of the DW transmission for different virtual source positions, $r_v$, ranging from 10.5 to 100 mm.

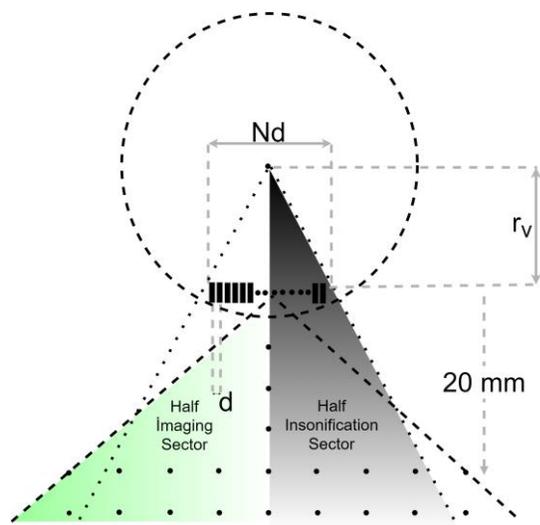

Fig. 5. The geometry used to derive phase profile for a particular DW.

We used the horizontal lines of pin targets to optimize the DW insonification for a particular field of view. For 90° imaging sector, the pin target line at 20 mm distance must be fully insonified. For a 60° sector, 5 targets at the center of the line at 20 mm depth and all targets of the line at 40 mm depth must be insonified. A 30° sector covers the 5 pin targets at the center of the line at 40 mm depth.

The signal strength of 8-chip coded DW transmission along the pin targets is shown in Fig 6. The pin targets are those encircled with dashed rectangle in Fig. 4, which are 25 mm away from the aperture. The maximum signal strength is achieved at the central pin target. The signal strength decreases for pin targets away from the center and 10 dB lower for the outermost targets. This difference is partly due to the effect of attenuation on increased path length at outermost pin targets. Other effects to consider are the wider point spread function and poorer insonification at larger sector angles. However, the signal-to-speckle ratio (SSR) of 15 dB at any pin target is preserved. When the virtual source distance, $r_v$, is set to about 14 mm or less, all pin targets positioned at 25 mm depth are detected with 15 dB SSR.

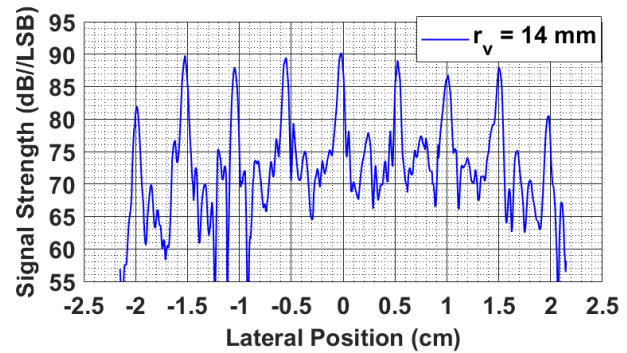

Fig. 6. Signal strength for 8-chip coded DW transmission when $r_v = 14$ mm along the horizontally spaced pin targets positioned at 25 mm depth.

The signal amplitude unit in this work is in terms of least significant bit (LSB) of DiPhAS acquisition system. The received signal amplitude is plotted in dB//LSB. When the speckle power is used as in Fig. 6, the squared signal amplitude is plotted in dB relative to the square of 1 LSB, i.e., 1, of signal. Therefore, both signal power and signal amplitude yield the same dB value and we refer to the vertical (signal) axis as signal strength.

To find the effect of virtual source position on the performance of coded DW transmission, we further investigated the signal strength variation across the same pin targets by varying $r_v$. The signal strength for different $r_v$ is shown in Fig. 7. The virtual source distance is increased from 10.5 to 21 mm. When $r_v$ is set to 21 mm, the transmitted energy is concentrated in a rather acute sector, which can be observed as increased signal strength of the center pin targets. However, it also causes reduction in signal strength by about 20 dB at larger angles. To spread the transmitted energy within the entire region of interest, the virtual source has to be closer to the array. When $r_v$ is 10.5 mm, the signal strength at the center pins is 2-3 dB lower but the difference between center and outermost targets is reduced to 4dB.

The entire sector of interest must be adequately insonified while the available energy in DW must be confined in the sector as much as possible. We used the signal strength distribution across the sector obtained in STA imaging as the reference to determine the correct virtual source distance for a given aperture size and field of view.

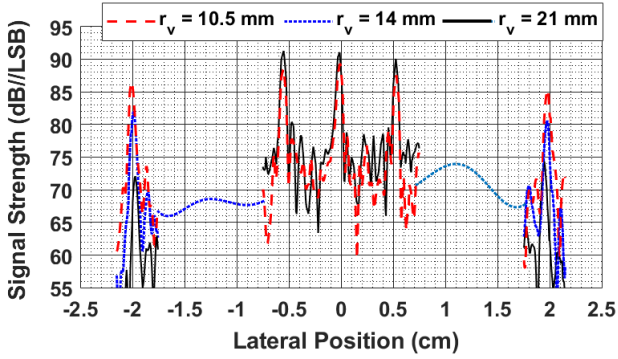

Fig. 7. The signal strength of 8-chip coded DW, as a function of virtual source position, along the horizontally spaced pin targets positioned at 25 mm depth. $r_v$ is increased from 10.5 to 21 mm.

### B. Beamforming for STA Imaging

STA imaging technique provides the highest possible signal strength. Therefore, we compared the DW imaging with STA imaging in terms of signal strength variation along the horizontally spaced pin targets.

We used 2-cycle pulse for STA imaging. A single array element is used at a time for transmission and all elements are used for reception. The measurement for STA imaging is repeated until all array elements are used for transmission. The recorded data for each transmission event is then correlated with the reference signal in correlation receiver. The delay and sum beamforming is applied on the correlator output as described in [34]. The beamformed signals obtained for each transmission event are summed up. The beamformer output is then envelope detected, log compressed and scan converted to form the entire image.

### C. Signal Strength Distribution Along Horizontally Spaced Pin Targets

The measurement results in terms of signal strength of STA and DW imaging techniques along the horizontally spaced pin targets are shown in Fig. 8. The pin targets are positioned at 25 mm and 40 mm depth, which are encircled with dashed and solid lines, respectively, in Fig. 4. The signal strength variation of DW becomes similar to that of STA when $r_v$ is 14 mm as illustrated in Fig. 8 at 25 mm and 40 mm depth. Furthermore, the signal strength of coded DW transmission outperforms that of the STA by more than 3 dB. In other words, signal gain in perfect transmit focusing of STA scheme is counter balanced and exceeded by 8-bit Golay coded DW transmission as far as signal strength is concerned.

Since this DW scheme requires only two transmissions for an image frame, 5000 fps is possible with signal strength better than that can be achieved by STA imaging.

14 mm virtual source distance and 12.8 mm array size (64λ at 7.5 MHz) subtends 49.1° for the insonification sector, which can cover only 31.1 mm lateral sector width at 20 mm distance from the array and not 40 mm as required by 90° imaging. The discrepancy between the geometrical prediction and the actual (achieved) coverage is due to diffraction. Limiting the insonification sector to the required imaging sector considering the diffraction effects improves the signal strength significantly.

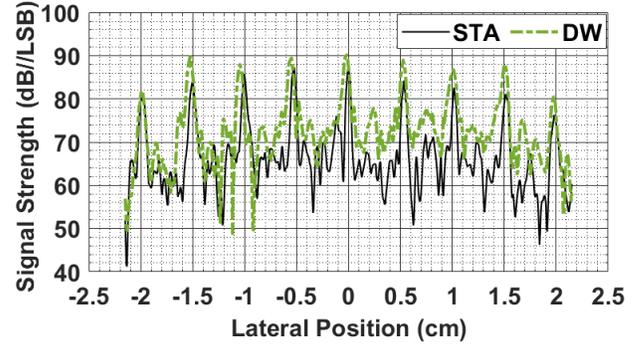

(a)

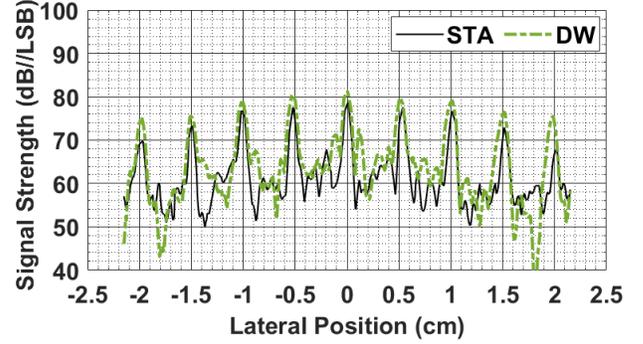

(b)

Fig. 8. Signal strength for 8-chip coded DW transmission when $r_v$ =14 mm and STA imaging with 1 chip pulse along the pin targets located at (a) 25mm depth (b) 40 mm depth.

### D. Optimizing the Virtual Source Location for Aperture Size and Sector Angle

We investigated the optimum virtual source position for different aperture sizes and sector angles. We used the horizontal lines of pin targets to optimize the virtual source position. We used 8-bit Golay coded DW transmission. The received signal strength achieved by coded DW transmission is compared with STA imaging technique. Table I presents the optimal virtual source position calculated for 64 λ and 32 λ aperture sizes at 90°, 60°, and 30° sector angles. The signal strength difference between the coded DW and STA imaging techniques are also presented. In order to generalize the results, we expressed the aperture size and virtual source distance in terms of λ at the center frequency of 7.5 MHz, in the Table I. The signal strength difference for 90° sector is given for the central pin target positioned at 25 mm (125 λ at 7.5 MHz) depth. For 60° and 30° sector angles, it is given for the central pin target at 20 mm (100 λ) depth.

Long virtual source distances are required for optimum DW delay profile as the sector angle decreases from 90° to 30° both in 64 λ and 32 λ aperture sizes. The decreased sector angle enables us to confine the energy in a rather acute sector. The delay profile of 32 λ aperture size must be configured for short virtual source distance compared to 64 λ. It is due to the fact that small array aperture requires a larger beam width to illuminate the entire region of interest. The signal strength of 8-bit Golay coded DW transmission exceeds STA imaging in all aperture lengths and sector angle combinations. For example, the signal strength obtained by coded DW transmission at the central pin target is 10.1 dB higher for 30° sector at 20 mm depth.

TABLE I
OPTIMAL VIRTUAL SOURCE POSITION

| Aperture Size | Sector Angle | $r_v$ | Received Signal Strength Difference (dB) |
|---|---|---|---|
| 64 λ | 90° | 70 λ | 3.1 |
| 64 λ | 60° | 150 λ | 6.3 |
| 64 λ | 30° | 250 λ | 6.1 |
| 32 λ | 90° | 26 λ | 7.1 |
| 32 λ | 60° | 40 λ | 8.5 |
| 32 λ | 30° | 70 λ | 10.1 |

## IV. Performance Of Coded DW Transmission

### A. SNR Comparison

SNR is typically defined as the ratio of signal power to the noise power and expressed in dB [35]. In medical ultrasound imaging, SNR is regarded as speckle to noise ratio and measured as $SNR_{+1}$ given in (2) [36]. Since the received signal already contains noise,

$$SNR_{+1} = 10 \log_{10} \left( \frac{P_{speckle}}{P_{noise}} + 1 \right) \qquad (2)$$

where $P_{speckle}$ is the speckle power and $P_{noise}$ is the noise power. $SNR_{+1}$ is similar to SNR only for large values. The speckle power is equal to noise power when $SNR_{+1}$ is 3 dB.

We measured the SNR for two different image items, the speckle and the pin targets. We used the focal zone of CSF phased array imaging as a reference for SNR comparison. The CSF phased array imaging provides constructive addition of the pressure signals in the focal zone. Thus, it yields high resolution and maximum achievable SNR in the focal region.

*1) Beamforming for CSF Phased Array Imaging:* In CSF phased array imaging measurements, we transmitted 2-cycle pulse. 181 steered and focused beam transmissions are used to construct the image scan lines sequentially. The steering angle ranges from -45° to 45° with respect to normal and spaced at 0.5° intervals. Each transmission is focused at 40 mm away from the transducer array center. All array elements are used in each transmit and receive event.

We applied receive beamforming directly on the raw data at each array element to form a scan line along the beam axis. A gaussian filter is then applied on the beamformed signal to reduce the high frequency noise components prior to envelope detection. The construction of entire image typically relies on receive delay and sum beamforming followed by envelope detection, log compression and scan conversion.

*2) SNR Comparison within Speckle Region:* We measured the SNR in speckle region (no targets), which is cross-hatched in Fig. 4. The SNR for coded DW transmission with different code lengths when $r_v$ is 14 mm (70λ) is compared with CSF imaging.

Single pulse and 2-to-10-bit CGS coded signals are transmitted into the speckle region. We also recorded the noise in each channel similar to coded signals, but without any transmission, and processed each channel noise exactly similar to the received coded signals. The array is kept acoustically in contact with the phantom surface to ensure the noise contribution of radiation resistance.

The $SNR_{+1}$ is shown in Fig. 9a as a function of depth. The SNR is improved by 3 dB up to 4.5 cm when the code length is increased by two folds, from 2 to 4 and 4 to 8 bits. In fact, when the code length is doubled, the speckle strength at the correlator output is increased by 6 dB whereas the noise strength is increased by 3 dB as shown in Fig. 9b. The SNR is improved by approximately 6 dB up to 4.5 cm depth in 2-bit coded signal compared to pulsed transmission. Additional 3 dB in SNR is due to the two transmissions in 2-bit coded signal. However, the gain in SNR achieved by increased code length decreases beyond 4.5 cm depth.

The noise strength is obtained by averaging the squared signals of independently taken 13 noise measurements. The significant difference between Gaussian filter and MF noise is mainly due to the difference in the frequency response of the respective filters.

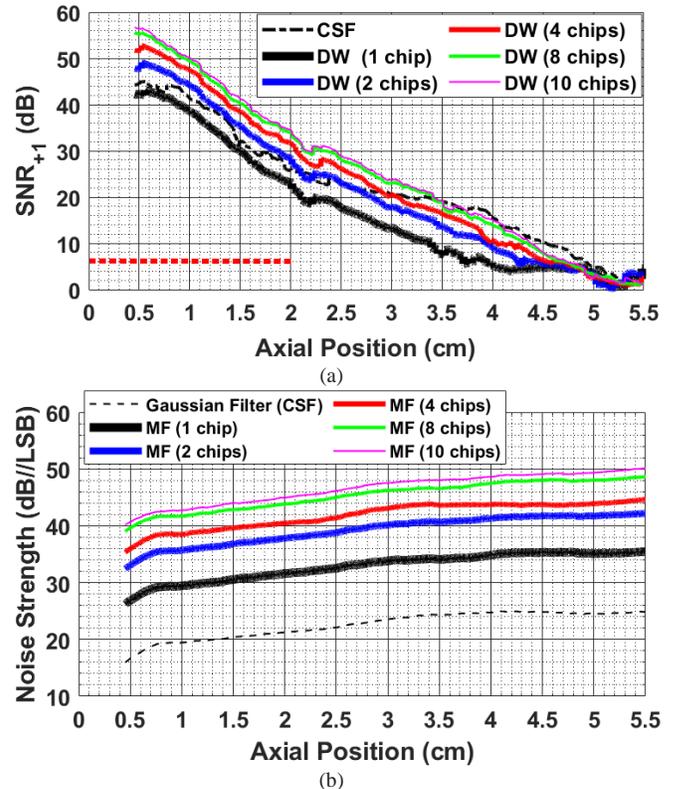

Fig. 9. Speckle region measurement results for CSF phased array and coded DW ($r_v$ =14) imaging techniques. (a) $SNR_{+1}$ (b) Noise Strength.

The SNR of CSF phased array imaging is maximum at 4 cm depth, which is the transmit focus. The SNR of 8-chip coded DW signal is 2 dB lower than that of the CSF phased array imaging in the focal region. However, the SNR of CSF phased array imaging deteriorates in non-focal regions. For example, the SNR of 8-bit Golay coded signal is 8.2 dB larger than that of the CSF phased array imaging at 2 cm depth.

Penetration depth can be determined from SNR measurements. It is defined as the depth at which $SNR_{+1}$ falls below 6 dB [36]. The 6 dB reference line is shown with dotted line in Fig. 9a. The CSF phased array imaging offers largest penetration depth, 5 cm, among all schemes. 5 cm is very close to the focus of the CSF transmission. The penetration depth of 8-bit Golay coded DW transmission is 4.75 cm, which is very close to that of the CSF phased array imaging.

*3) SNR for Pin Targets:* We also measured the SNR at pin targets. The SNR along the horizontally spaced pin targets positioned at 40 mm depth and vertically spaced pin targets are shown in Fig. 10, respectively. The SNR of 8-chip coded DW signal at the horizontally spaced pin targets is 4 dB less than that of the CSF phased array imaging in the focal region (center pin) and they are equal at the outermost pins, when $r_v$ is 14 mm. If MF is used instead of Gaussian filter in CSF phased array imaging, the SNR is further improved by 4 dB. It is clearly seen in Fig. 10b that the coded DW has significant SNR advantage over the CSF phased array imaging in the non-focal region. CSF imaging with MF maintains up to 4 dB SNR advantage beyond the focal region. A ninety-fold increase in the frame rate is achieved because there are two transmissions in coded DW scheme compared to the 181 transmissions in conventional method.

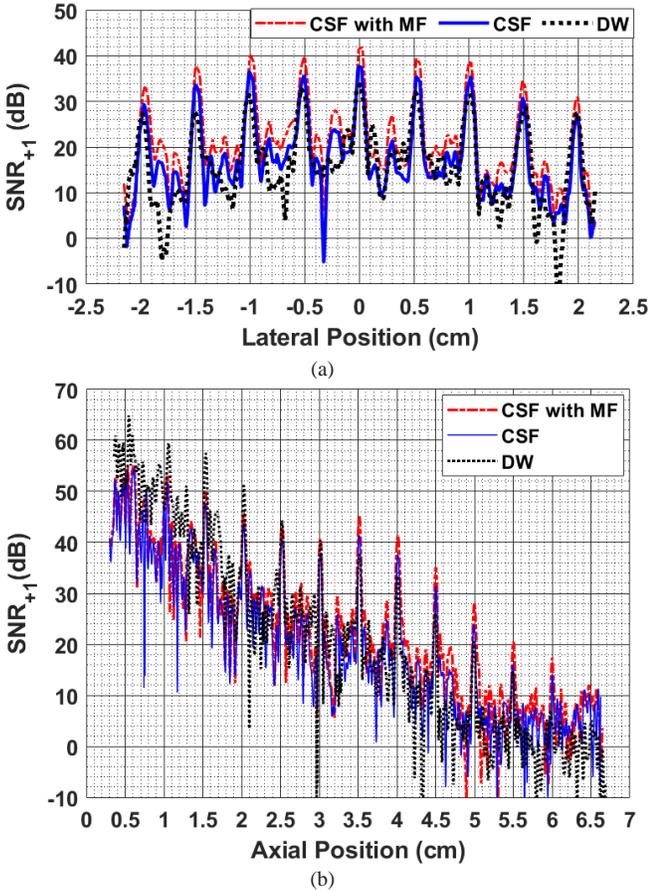

Fig. 10. SNR for CSF phased array imaging and coded DW imaging (8 chips and $r_v$ =14 mm). CSF phased array imaging performed with Gaussian filter and MF, respectively, is shown for comparison. (a) SNR for horizontally spaced pin targets at 40 mm depth (b) SNR for vertically spaced pin targets.

## B. CNR Comparison

We also used contrast-to-acoustic noise (speckle) ratio (CNR) to evaluate the performance of coded DW scheme. The CNR is a quantitative measure to determine the contrast in the ultrasound image. It is defined in terms of the mean and variance in the image of lesion region and the background speckle. We calculated CNR by using the following expression [34], [36]:

$$CNR = \frac{|\mu_C - \mu_B|}{\sqrt{\sigma_C^2 + \sigma_B^2}} \quad (3)$$

where $\mu_C, \mu_B, \sigma_C$ and $\sigma_B$ are the means and standard deviations of the log image pixel values of the cyst and background regions, respectively. CNR for anechoic cyst structure, which is encircled with dotted rectangle in Fig 4, is investigated. The cyst has 4 mm diameter and is positioned at 40 mm depth, the focal distance of the CSF image. CNR was calculated in the circular regions indicated with white circles in Fig. 11. The variation of CNR with respect to circular region diameter for CSF and DW imaging techniques is depicted in Fig. 11a.

CNR calculation yields different results in different frames in coded DW transmission, since the number of statistically independent pixels contributing to mean and variance of CNR calculation is limited in a single frame. We used 8-bit Golay coded DW transmission when $r_v$ is 14 mm. 11 transmissions were performed at different transducer array positions, in order to have statistically independent data suitable to assess the frame-to-frame variation of CNR. The transducer array is moved across the phantom surface at approximately 1 mm steps. The variation of CNR in two different frames, yielding minimum and maximum CNR, is plotted in Fig. 11a. CNR in coded DW images are lower than the CNR obtained in CSF image. CNR obtained in CSF image is 2.75 when circular region diameter is 2 mm whereas the CNR for one frame of the coded DW transmission varies between 0.44 and 2.04, with 1.5 mean and 0.43 standard deviation.

The cyst image obtained by CSF phased array imaging is shown in Fig. 11b in which the CNR is 2.75. The DW images with 2.04 CNR and 0.44 CNR are given in Fig. 11c and 11d, respectively.

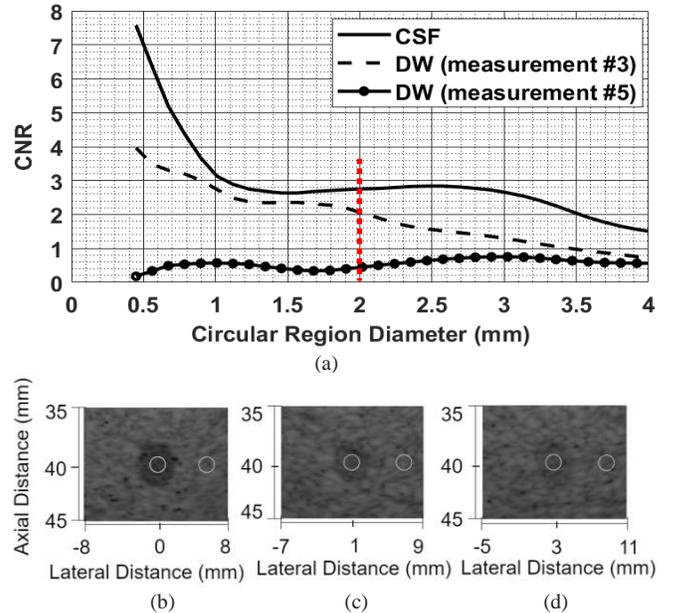

Fig. 11. CNR for anechoic cyst with 4 mm diameter positioned at 40 mm depth: (a) CNR variation against evaluation region size for CSF image and for DW images yielding maximum and minimum CNR. The cyst images obtained by (b) CSF phased array imaging (c) Coded DW imaging. The transducer is moved by 1 mm. (d) Coded DW imaging. The transducer is moved by 3mm.

## V. CODED DW IMAGE

The ultrasound images of the pin targets are presented in Fig. 12. The dynamic range of all images is set to 60 dB and the local mean of all images is set to 32 dB for brightness uniformity.

CSF phased array imaging focuses the transmitted pulses at the transmit focusing point. It offers high resolution and SNR at 40 mm depth, where the pin targets are easily distinguishable, as shown in Fig 12a. In unfocused region, the transmitted signals are not coherently superimposed and SSR is decreased. This decrease is apparent at the pin targets located at 20 mm depth or less. 55 fps can be achieved in CSF imaging, when 181 transmissions per frame is used to construct the image. In STA imaging, there is a gain in signal strength at pin targets due to perfect focusing, but also loss in SNR due to 128 times more noise power compared to CSF phased array imaging. STA image is given in Fig. 12b. The pin targets at deeper regions beyond 45 mm depth are not detected because of poor SNR. The frame rate in STA depends on the number of transmit elements. Since 128 elements are used, 78 fps is possible.

The image of one frame obtained by 8-bit coded DW transmission when $r_v$ is 14 mm is shown in Fig. 12c. This image is obtained using only two transmissions. Coded DW transmission allows to construct an image at 5000 fps rate. All pin targets from 5 mm to 55 mm are clearly defined in the image and the farthest target in 60 mm is visible. Similarly, horizontal targets at 20 and 25 mm are well defined. The images of outermost targets at 40 and 45 mm are slightly smeared out.

The clarity and definition in the images obtained by the three imaging techniques are in agreement with the quantitative analysis results.

## VI. DISCUSSION OF RESULTS

### A. Performance Variation

In this study, we optimize the use of coded signals in DW transmission to maximize the SNR while covering the entire sector with a received signal strength distribution equivalent to STA imaging. The overall performance of coded DW transmission was quantified for one frame obtained at very high frame rate. The performance metrics are SNR, CNR, speckle power and target signal strength. We present the performance variation as a function of code length, wave profile, transmit/receive aperture size and field of view.

Coded DW imaging offers signal strength variation across the viewing sector similar to STA imaging when $r_v$ is 14 mm. 8-chip coded DW imaging provides approximately 90-fold increase in frame rate without compromising from SNR compared to CSF phased array imaging even in the focal regions. It also offers increased SNR in other regions. 5000 fps rate is possible at the imaging depth of 7.5 cm by using coded DW transmission.

The CNR in coded DW scheme is lower, when 8-bit coded DW and CSF imaging are compared on a single frame basis. The variation of CNR from frame to frame is significant in coded DW imaging. The mean CNR is 1.5 with 0.44 standard deviation in 11 images, each obtained in 200 microseconds using 8-bit coded DW transmissions. CNR obtained for the same cystic region in CSF imaging is 2.75 at the focus.

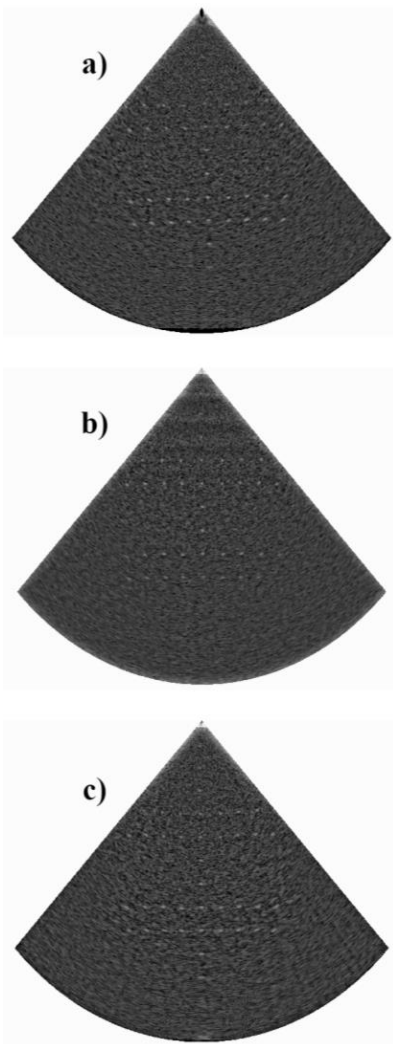

Fig. 12. Ultrasound images of the pin targets. (a) CSF phased array imaging (b) STA imaging (c) Coded DW imaging.

### B. Limitations and Potential Improvements

The performance of coded transmission relies on maintaining high correlation between the received coded signals and the reference signal. We found out that using attenuation compensated reference signal significantly improves the SNR. Refinement in the attenuation compensation of the reference signal will further improve the SNR.

Ultrasound imaging is a very wideband application in highly attenuating media whereas the conceptual infrastructure of coded signals is developed for applications in comparatively narrow band and rather lossless media. Research for short-duration chip waveforms which maintain better correlation properties in attenuating medium will improve the performance of coded ultrasonic imaging.

We employed CGS codes in this work, because these codes do not have any code range lobes. However, the self-interference caused by incoherent addition of signals in DW transmission and the deterioration in correlation properties of the coded signals due to attenuation can be more significant than code range lobes. In this respect, use of other codes, like gold sequences, m-sequences, etc., must also be investigated to find out if similar performance can be obtained with single transmission. Also, we did not use apodization in this work.

Determination and use of appropriate apodization for coded DW imaging are also a potential area for improvement.